\documentclass[11pt]{article}
\usepackage{hyperref}
\usepackage{graphicx}
\pdfoutput=1
\begin{document}
\title{Megahertz Schlieren Imaging of Shock Structure and Sound Waves in Under-Expanded, Impinging Jets}
\author{C. Willert$^1$, D. Mitchell$^2$, J. Soria$^2$ \\[9pt]
$^1$) Institute of Propulsion Technology,\\
    German Aerospace Center (DLR),
    51170 K\"oln, Germany\\[6pt]
$^2$) Mechanical and Aerospace Engineering, Monash University\\
    Melbourne, VIC 3800, Australia}
\maketitle
%

\begin{abstract}

The accompanying fluid dynamics videos visualize the temporal evolution of shock structures and sound waves in and around an under-expanded jet that is impinging on a rigid surface at varying pressure ratios. The recordings were obtained at frame rates of 500~kHz to 1~Mhz using a novel pulsed illumination source based on a high power light emitting diode (LED) which is operated in pulsed current mode synchronized to the camera frame rate.
\end{abstract}


\section{Introduction}

The high speed Schlieren visualizations  have been made possible by recent advances in light emitting diode (LED) technology that has resulted in high power, single chip devices with luminous radiant fluxes exceeding several watts. When operated in pulsed high current mode the instantaneous light emission of an LED can be increased significantly \cite{Stasicki:1984,Hiller:1987}. By overdriving modern high power LEDs intensity levels are reached that comparable to the intensity of photographic (xenon) flash units and hence are a suitable light source for Schlieren imaging \cite{Willert:2010}. Compared to the commonly used xenon flash units an LED can be triggered within tens of nanoseconds at rise times on the order of 100~ns thereby enabling stroboscopic illumination at megahertz rates.

In the present application the LED's driving electronics were synchronized to a high speed camera (Shimadzu HPV-1, $312 \times 260$ pixels) to provide time-resolved Schlieren images of an underexpanded free jet impinging on a flat plate at nozzle pressure ratios from 2.0 to 5.0. The LED (Luminus, Phlatlight CBT-120, green \cite{datasheet:cbt120}) was pulsed in burst mode for 102 images at currents of up to 120 Ampere with 500~ns pulse duration. Due to the finite rise time provided by the LED drive electronics (200 ns) the actual light pulses had a  duration of roughly 300~ns. Compared to images obtained with a xenon white light flash the nearly monochromatic green light of the LED results in much crisper flow features with superior repeatability in intensity without any speckle artifacts commonly found with laser illumination.

\begin{figure}
\includegraphics*[width=.78\textwidth]{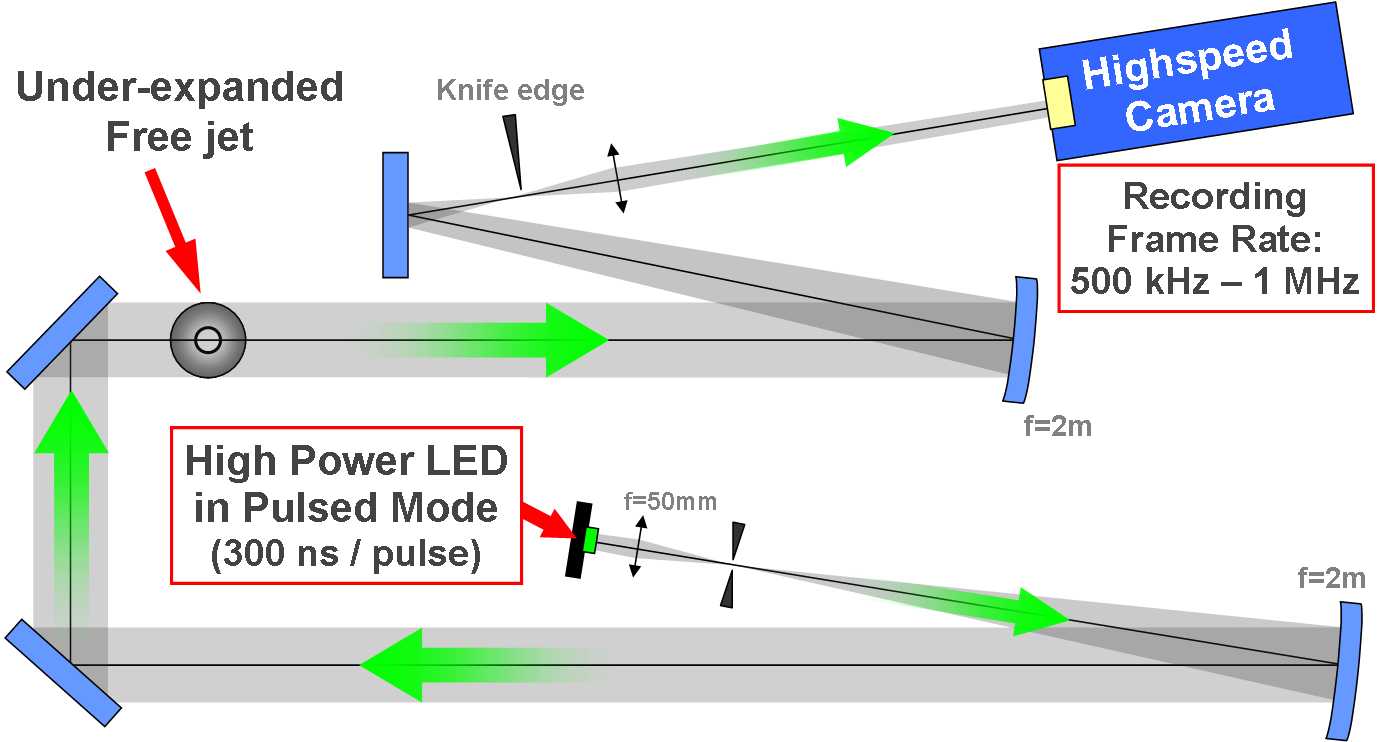}
\caption[]{Schlieren imaging setup with pulsed high power LED as light source)\label{fig:schlieren_setup}}
\end{figure}

The fluid dynamics video 
shows four sequences of the unsteady impinging jet flow at a fixed plate distance of $x/D = 4$, with $D=5$~mm being the nozzle diameter. The respective nozzle pressure ratios are: 2.0, 2.5, 3.0 and 5.0. The details of the jet facility and the schlieren setup are described in \cite{Risborg+Soria-2008, Risborg-5ACLDFMC-2008}. A schematic of the setup is shown in figure~\ref{fig:schlieren_setup}.

Each sequence of the fluid dynamics video shows the original schlieren recording on the lower half and simultaneously a contrast enhanced, pseudo-colored version on the upper half. Contrast enhancement was achieved by subtracting the average image from the originals which has the effect of highlighting only the unsteady features in the images. Static features such as blemishes on the mirror surfaces are removed.

The schlieren flow visualization show a distinct change in the structure of the supersonic under-expanded jet as the pressure ratio increases. At the lower pressure ratios of 2.0 and 2.5 the weak shocks are located closer to the jet exit and are observed to gyrate, suggesting a helical motion of the entire jet most likely due to a helical instability.  Shear layer instabilities are clearly observable on the parallel jet boundary emanating from the convergent nozzle.  This jet boundary on which the shear layer instabilities ride is observed to become bowed at the higher pressure ratios of 3.0 and 5.0 with the curvature increasing with pressure ratio. At pressure ratios above 3.0 the formation of a Mach disk at the core of the jet is evident.  The Mach discs do not gyrate at these higher pressure ratios, however, a highly turbulent flow is observed downstream of the Mach discs which are separated by a larger distance. A slip line forms around the edges of the Mach disk, resulting in the generation of internal shear layers, and thus the increase in turbulence intensity immediately downstream of the shock. In all cases a turbulent boundary layer flow developed on the impinging plate and acoustic waves are observed to radiate away from the impingement zone. The sound waves radiate asymmetrically from this source region at the lower pressure ratios of 2.0 and 2.5, most likely due to the helical jet structure observed in these cases and hence, asymmetric impingement on the flat plate resulting in an asymmetric acoustic sound source. In the higher pressure ratio cases the sound waves radiated are more clearly observable suggesting higher acoustic pressures and in these two cases the radiated sound waves from the impingement zone have a more symmetric appearance, which is consistent with the more symmetric nature of the higher pressure ratio impinging supersonic under-expanded jet.

Beyond the application shown here pulsed LED illumination has been used as a light source for flow field measurement, in particular particle image velocimetry (PIV) \cite{Willert:2010,Stasicki:2010}, as well as in spray diagnostics \cite{ILASS:2010}.

\bibliographystyle{cew_bib}
\bibliography{Biblio_CWillert}

\end{document}